\documentclass[aps,twocolumn,pra,superscriptaddress,showpacs,tightenlines]{revtex4}
\usepackage{amssymb}
\usepackage{amsmath}
\usepackage{graphicx}
\usepackage{epsfig}
\usepackage{subfigure}
\usepackage{amsfonts}

\input{tcilatex}
\begin{document}

\title{Analog of Electromagnetically Induced Transparency Effect for Two Nano/Micro-mechanical Resonators Coupled With Spin Ensemble}
\author{Yue Chang}
\author{C. P. Sun}
\email{suncp@itp.ac.cn} \homepage{http://www.itp.ac.cn/~suncp}
\affiliation{Institute of Theoretical Physics, The Chinese Academy
of Sciences, Beijing, 100080, China}

\begin{abstract}
We study a hybrid nano-mechanical system coupled to a spin ensemble
as a quantum simulator to favor a quantum interference effect, the
electromagnetically induced transparency (EIT). This system consists
of two nano-mechanical resonators (NAMRs), each of which coupled to
a nuclear spin ensemble. It could be regarded as a crucial element
in the quantum network of NAMR arrays coupled to spin ensembles.
Here, the nuclear spin ensembles behave as a long-lived transducer
to store and transfer the NAMRs' quantum information. This system
shows the analog of EIT effect under the driving of a probe
microwave field. The double-EIT phenomenon emerges in the large $N$
(the number of the nuclei) limit with low excitation approximation,
because the interactions between the spin ensemble and the two NAMRs
are reduced to the coupling of three harmonic oscillators.
Furthermore, the group velocity is reduced in the two absorption
windows.

\end{abstract}

 \maketitle


\section{Introduction}

In quantum information, an important task is the long-lived storage and
remote quantum state transfer~\cite{inf,cirac,bennet,shi1,shi2} of quantum
information. There exist several approaches to the implementation of quantum
storage, such as electromagnetically induced transparency (EIT) based on
three-level atomic ensemble~\cite%
{harris,hau,scully,lukin,lukin1,lukin2,hau1,liyong,liyong1}, nuclear spins
coupled to electrons~\cite{zhangp}, and polarized molecular ensembles
coupled to cavity fields in superconducting transmission lines~\cite%
{zhou,liao1,liao2}. Nuclear spin ensemble has the advantage that its
transverse relaxation time $T_{2}$ can reach a second time scale~\cite%
{lukin3,t2}. In earlier works~\cite{lukin3,zoller2,poggio1,zhangp}, the
nuclei ensemble has been used to store the quantum information of electron
spins, since the electron spin's decoherence time $T_{e2}$ is in the order
of ten milliseconds~\cite{t2,te2}, which is much shorter than the nuclear
spins' relaxation time.

Recently, the optomechanical systems containing micro/nano-mechanical
resonators have inspired extensive studies in many aspects, such as the
entanglements of the mechanical resonators with the light~\cite%
{mancini,knight,penrose,vitali}, and even the atoms~\cite{vitali1,ian,chang}%
, cooling the mechanical resonators through light pressure~\cite%
{meystre0,meystre,meystre1,vitali2,zoller0}, and the nonclassical states in
the hybrid system~\cite{knight1,gong,zoller1}. In fact, the
micro/nano-mechanical resonator's decoherence time $T_{r2}$ is shorter~\cite%
{tr20,tr2} ($\sim $100 $\mu $s) than the life time of the nuclear spins.
Therefore, it is expect to store the information of the
micro/nano-mechanical resonator in the nuclear spins. Actually, the coupling
between the nuclear spin ensemble (or a single spin) and the mechanical
resonator tips has drawn much attention~\cite%
{rugar,sidles,rugar1,rugar2,rugar3,rugar4,rugar5,xuef} both in theories and
in experiments. An important innovation based on the coupling of single/few
spins to the mechanical tip is the magnetic resonance force microscopy
(MRFM)~\cite{rugar,sidles,rugar1,rugar2,rugar3,rugar4,rugar5,poggio}. MRFM
uses a cantilever tipped with a ferromagnetic particle producing a
inhomogeneous magnetic field that couples the mechanical tip to the sample
spins. By measuring the displacement of the tip with an interferometer, a
series of 2-D images of the spin sample is acquired~\cite{mri}. In practice,
the spin sample is usually a spin ensemble containing a lot of electrons or
nuclear spins, which could be excited to show the collective behavior. Such
collective motion could achieve the effective strong coupling to the
nano-mechanical resonator (NAMR).

\begin{figure}[tbp]
\includegraphics[bb=28 458 585 783, width=8 cm, clip]{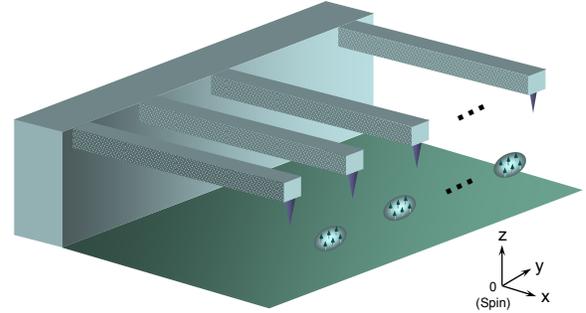}
\caption{(Color online) Schematic setup of the NAMR-spin-NAMR-...
system. Here, the spin ensembles are placed between two nearest
NAMRs in the NAMR array, and each of the NAMR has a tiny
ferromagnetic particle in the mechanical tip. The spin ensembles
behave as a transducer that store the NAMRs' quantum information and
transfer them from one NAMR to the next one.}
\end{figure}

With the above mentioned investigations about various hybrid systems
concerning the nuclear spin ensembles and NAMRs, Rabl et.al.~\cite{lukin4}
explore the possibility of using the short life time NAMR as a quantum data
bus for spin quibit coupled to magnetized mechanical tips, and the
mechanical resonators are coupled through Coulomb forces. This study
motivates us to utilize the nuclear spin ensemble itself as long-lived data
bus (the spin ensemble also behaves as a quantum transducer~\cite{liu}) to
realize the effective couplings among the NAMRs. The advantage of our
proposal is that the quantum transducer has the life time much longer than
the NAMR's. Our setup is shown in Fig. 1, where an array of NAMRs is coupled
to nuclear spin ensembles, which are placed between the nearest two tips.
Each spin ensemble induces interaction between the corresponding tips, and
the quantum information of the tips can be transferred from one to the
another one by one. This dynamic process realizing the quantum information
transfer physically depends on an controllable coupling among the three
systems, two NAMRs and a spin ensemble. We will show that the double EIT
effect exists in our present setup, which plays an important role in the
coherent storage of quantum information in this hybrid-element sub-system.

In the conventional EIT effect based on the $\Lambda $-type three-level
atomic ensemble on two-photon resonance, a driving light suppresses the
absorption of another light (the probe light), and even makes the probe
light transparent at the frequency at which the probe light should be
absorbed strongly without the driving field~\cite{scully1}. An important
physical mechanism in this EIT effect is that the pump light induces an
ac-Stark splitting of the excited state. As a result, the probe light is
off-resonant with the energy spacing of the energy levels it couples to.
Actually, the EIT effect analog exists in a system of two coupled harmonic
oscillators one of which is subject to a harmonic driving force~\cite{alzar}%
. In fact, the coupling between the two harmonic oscillators will change
their original frequencies, and make the absorbed power deviate from
resonance. This reason is similar to that in the conventional EIT
phenomenon.\ We show that our proposed setup consisting of a magnetized
mechanical tip coupled to a nuclear ensemble, which behaves as a two coupled
harmonic oscillator system, can also exhibit the phenomenon similar to the
EIT effect in the system with light-atom interaction.

We will study in details the double EIT effect analog in a sub-network of
the whole structure shown in Fig. 1, a NAMR-spin ensemble-NAMR coupling
system. In the low excitation limit with large $N$ (the number of the
nuclear spins) limit, the spin excitation behaves as a single mode boson~%
\cite{liyong1,zhangp,ian} coupled respectively to the two mechanical tips.
In this case, the interaction between the spins and each tip is the coupling
between two harmonic oscillators with effective amplified strength
proportional to $\sqrt{N}$. In general, this three oscillator-coupling
system have three eigen-frequencies (taking account of the degeneracy). And
we show that there are two absorption windows for the probe microwave field,
with the absorption peaks corresponding to the three eigen-frequencies. In
these two windows with normal dispersion relations, the group velocity of
the microwave field is reduced dramatically. These transparency and slow
light phenomena correspond to EIT effect.

The paper is organized as follows: in Sec. II, we illustrate the sub-network
composed of two nano-mechanical resonators coupled to a spin ensemble; in
Sec. III, we study the mechanical analog of EIT effect in a NAMR-nuclear
ensemble coupling system, and make a comparison with the AMO system by
revisiting the conventional EIT phenomenon; in Sec. IV, we study the
double-EIT effect in the sub-network hybrid system and show the slowing
light phenomenon in Sec. V; in Sec. VI, we summarize our result.

\section{Setup and Modeling for Quantum Transducer}

\begin{figure}[tbp]
\includegraphics[bb=43 370 584 783, width=8 cm, clip]{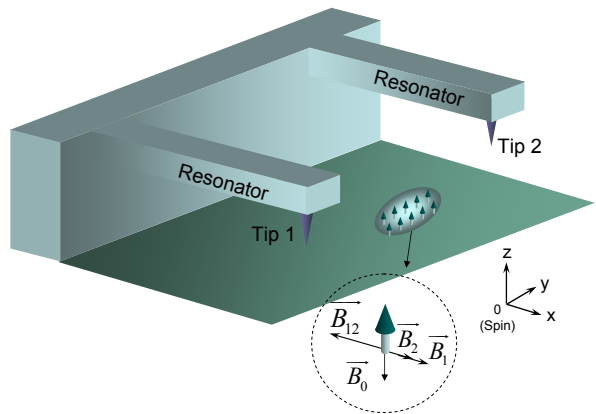}
\caption{(Color online) Schematic setup of the NAMR-spin ensemble-NAMR
coupling system. The ensemble of spins is placed between two NAMRs each of
which has a tiny ferromagnetic particle in the tip. Both of the directions
of the two magnetic field produced by the two tips are along the $x$-axis.
The origin of the coordinate frame is at the center of the spin ensemble.
The spin ensemble is also exposed in two static magnetic fields $\vec{B}%
_{12} $ along the $x$-axis, and $\vec{B}_{0}$ along the $z$-axis.}
\end{figure}

We now consider a hybrid system consisting of two NAMRs and a nuclear spin
ensemble containing $N$ spins. This system is the basic unit for
constructing the whole quantum network (Fig. 1). The spin-NAMR hybrid system
is illustrated in Fig. 2. In this setup, each NAMR is coupled to the
ensemble of $N$ $1/2$-spin particles by a tiny ferromagnetic particle
attached to it. The origin of the reference frame is chosen to be the center
of the nuclear spin ensemble. The NAMRs can oscillate in the $z$-direction,
and each magnetized tip attached to the corresponding NAMR produces a
dipolar magnetic field at the position of the spins as \cite{jackson}%
\begin{equation}
\vec{B}_{j}=\frac{\mu _{0}\left[ 3\left( \vec{m}_{j}\cdot \vec{n}_{j}\right)
\vec{n}_{j}-\vec{m}_{j}\right] }{4\pi r_{j}^{3}},\text{ }j=1,2,
\end{equation}%
where $\mu _{0}$ is the vacuum magnetic conductance, $\vec{m}_{j}$ is the $j$%
the ferromagnetic particle's magnetic moment, $\vec{n}_{j}$ is the
corresponding unit vector pointing in the direction from the tip to the
spin. Here, $r_{j}$, which varies due to the oscillation of the NAMR along
the $z$-direction, is the distance between the magnetic tip and the spin. In
our setup, both of the magnetic moments in the two tips are in the $x$%
-direction as $\vec{m}_{1}=m_{1}\hat{e}_{x}$ and $\vec{m}_{2}=m_{2}\hat{e}%
_{x}$. The equilibrium positions of the two NAMRs are $\vec{r}_{1}$ and $%
\vec{r}_{2}$ respectively, and both $\vec{r}_{1}$ and $\vec{r}_{2}$ are in
the $yz$-plane. We have assumed that the spins are confined in a very small
volume, and the magnetic fields produced by the two ferromagnetic particles
at the spin ensemble are uniform as $\vec{B}_{1}=\left( B_{1}\left(
z_{1}\right) ,0,0\right) $ and $\vec{B}_{2}=\left( B_{2}\left( z_{2}\right)
,0,0\right) $ respectively, where%
\begin{equation}
B_{j}\left( z_{j}\right) \approx A_{j}-G_{j}z_{j}\text{, }j=1,2\text{,}
\end{equation}%
with $z_{1}\left( z_{2}\right) $ the small deviation of the tip1 (tip2) from
the equilibrium position. Here, $A_{1}=-\mu _{0}m_{1}/\left( 4\pi \left\vert
\vec{r}_{1}\right\vert ^{3}\right) ,$ $A_{2}=-\mu _{0}m_{2}/\left( 4\pi
\left\vert \vec{r}_{2}\right\vert ^{3}\right) $, together with the magnetic
field gradients%
\begin{equation}
G_{1}=\frac{3r_{1z}\mu _{0}m_{2}}{4\pi \left\vert \vec{r}_{1}\right\vert ^{5}%
}\text{, }G_{2}=\frac{3r_{2z}\mu _{0}m_{2}}{4\pi \left\vert \vec{r}%
_{2}\right\vert ^{5}}\text{,}
\end{equation}%
where $r_{jz}=\vec{r}_{j}\cdot \hat{e}_{z}$, for $j=1,2$. Besides these two
magnetic fields, the spins are also exposed to two static magnetic fields $%
\vec{B}_{12}=\left( -A_{1}-A_{2},0,0\right) $, and $\vec{B}_{0}=-B_{0}\hat{e}%
_{z}$. We note that in experiments~\cite{rugar1,rugar2,mn}, the distance
between the magnetized tip and the nuclear ensemble is in the order of 100
nanometers, and the nuclear spin ensemble containing more than 100 nuclei is
attached in a quantum dot with the diameter in 10 nanometers length scale.
Thus the the magnetic field $B_{j}\left( z_{j}\right) $\ is approximately
homogeneous in the nuclear ensemble when $z_{j}$ is fixed.

Both of the NAMRs are described as harmonic oscillators with effective
masses $M_{j}$ and frequencies $\omega _{j}$. Then the Hamiltonian $H_{0}^{d}
$\ of this spin-NAMRs coupling system is%
\begin{eqnarray}
H_{0}^{d} &=&\frac{p_{1}^{2}}{2M_{1}}+\frac{p_{2}^{2}}{2M_{2}}+\frac{1}{2}%
M_{1}\omega _{1}^{2}z_{1}^{2}+\frac{1}{2}M_{2}\omega _{2}^{2}z_{2}^{2}
\notag \\
&&+\sum_{j=1}^{N}\left( g_{1}\sigma _{j}^{x}z_{1}+g_{2}\sigma
_{j}^{x}z_{2}+g_{0}\sigma _{j}^{z}\right) ,  \label{4}
\end{eqnarray}%
where $p_{j}$ is the momentum of the NAMR $j$, $\sigma _{x}$ and $\sigma _{y}
$ are Pauli matrixes describing the spin. Here, the spin-NAMR coupling
strength $g_{j}=g_{s}\mu _{B}G_{j}/2$ for $j=1,2$, where $g_{s}$ is the
g-factor of the spin, $\mu _{B}$ is the Bohr magneton, and $g_{0}=g_{s}\mu
_{B}B_{0}/2.$ Note that the the first order in the magnetic dipole-dipole
interaction%
\begin{equation}
H_{d-d}=\frac{\mu _{0}\left[ 3\left( \vec{m}_{1}\cdot \hat{e}_{12}\right)
\left( \vec{m}_{2}\cdot \hat{e}_{12}\right) -\vec{m}_{1}\cdot \vec{m}_{2}%
\right] }{4\pi \left\vert \vec{r}_{1}-\vec{r}_{2}\right\vert ^{3}}
\end{equation}%
vanishes in our model, where $\hat{e}_{12}$ is the unit vector pointing in
the direction from the tip1 to the tip2.

To see the analog of EIT effect, we apply a probe microwave field $\vec{B}_{%
\mathrm{p}}=-\hat{e}_{x}B_{\mathrm{p}}\cos \Omega t$ coupled to the spin
ensemble. This coupling is described by the interacting Hamiltonian%
\begin{equation}
H_{I}=\frac{1}{2}g_{s}\mu _{B}B_{\mathrm{p}}\cos \Omega
t\sum_{j=1}^{N}\sigma _{j}^{x}.  \label{r}
\end{equation}%
The probe alternating magnetic field is similar to the probe light in the $%
\Lambda $-type atomic ensemble. The total Hamiltonian $H^{d}=H_{0}^{d}+H_{I}$
depicts the sub-network illustrated in Fig. 2.

When $N$ is large and with the low excitations of the spins, the excitations
of the spins are described by two bosonic operators~\cite{liyong1,zhangp,ian}%
\begin{equation}
b=\frac{1}{\sqrt{N}}\sum_{j=1}^{N}\sigma _{j}^{-}
\end{equation}%
and its conjugate $b^{\dag }$, where the commutation relation between $b$
and $b^{\dag }$ is%
\begin{equation}
\left[ b,b^{\dag }\right] \approx 1.
\end{equation}%
In terms of $b$ and $b^{\dag }$ defined above, the Hamiltonian in Eq. (\ref%
{4}) is rewritten as%
\begin{eqnarray}
H_{0}^{d} &=&\frac{\hbar \omega _{1}}{2}\left( P_{1}^{2}+Z_{1}^{2}\right) +%
\frac{\hbar \omega _{2}}{2}\left( P_{2}^{2}+Z_{2}^{2}\right)   \notag \\
&&+\frac{\hbar \omega _{0}}{2}\left( P_{0}^{2}+Z_{0}^{2}\right) +\hbar \sqrt{%
N}\sum_{j=1}^{2}G_{j}Z_{0}Z_{j}.
\end{eqnarray}%
Here, we have defined the dimensionless operators%
\begin{equation}
Z_{0}=\frac{b+b^{\dag }}{\sqrt{2}},P_{0}=i\frac{b^{\dag }-b}{\sqrt{2}},
\end{equation}%
\begin{equation}
Z_{j}=\sqrt{\frac{M_{j}\omega _{j}}{\hbar }}z_{j},P_{j}=\frac{p_{j}}{\sqrt{%
\hbar M_{j}\omega _{j}}},j=1,2.
\end{equation}%
The coupling constants $G_{j}=g_{j}\sqrt{2\hbar /M_{j}\omega _{j}}/\hbar ,$
for $j=1,2$, and $\omega _{0}=2g_{0}/\hbar .$ In experiments, the parameters
$\omega _{j}$ ($j=1,2$) and $\omega _{0}$ are in the order of 10$^{6}$Hz, and%
$\ G_{j}\ $can reach the order of $10^{5}Hz$.

Before the further investigations of the double-EIT effect in this hybrid
system, we would like to show the mechanical analog of EIT phenomenon in a
NAMR-spin ensemble coupling system with only a single NAMR, as the basic
physics in the double-EIT phenomenon depends on the coherent coupling of the
NAMR to the nuclear spin ensemble.

\section{Mechanical Analog of EIT}

In this section, we show the analog of the EIT effect in the single NAMR
coupled to a spin ensemble system. To this end, we will compare it with the
EIT phenomenon in the AMO system.

\begin{figure}[tbp]
\includegraphics[bb=32 482 545 789, width=8 cm, clip]{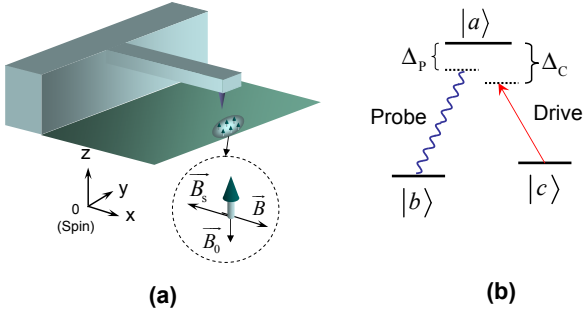}
\caption{(Color online) The schematic of a NAMR couple to a spin
ensemble [for (a)], and a $\Lambda $-type three-level atom [for
(b)]. The EIT effect base on the atomic ensemble [for (b)] where
each atom is coupled to a driving light and probe light has a analog
in the two harmonic oscillator coupling system derived from the
structure [for (a)]. In Fig (a), the spin ensemble is placed under
the NAMR which has a tiny ferromagnetic particle in the tip. The
direction of the magnetic field produced by the magnetized tip is
along the $x$-axis. The origin of the coordinate frame is at the
center of the spin ensemble. The spin ensemble is also exposed in
two
static magnetic fields $\vec{B}_{\mathrm{s}}$ along the $x$-axis, and $\vec{B%
}_{0}$ along the $z$-axis.}
\end{figure}

To reveal the basic physical mechanism, we first consider a system
consisting of a nano-mechanical resonator (NAMR) and a nuclear spin ensemble
containing $N$ spins. The spin-NAMR hybrid system is illustrated in Fig.
3(a). The origin of the reference frame is chosen to be the center of the
nuclear spin ensemble. The NAMR can oscillate along the $z$-direction, and
the magnetized tip attached to the NAMR produces a dipolar magnetic field at
the position of the spin, with the magnetic field $\vec{B}=\left( B\left(
z\right) ,0,0\right) $, where%
\begin{equation}
B\left( z\right) \approx A-Gz\text{,}
\end{equation}%
with $A=-\mu _{0}m/\left( 4\pi \left\vert \vec{r}\right\vert ^{3}\right) $,
and the magnetic field gradient is $G=3r_{z}\mu _{0}m/\left( 4\pi \left\vert
\vec{r}\right\vert ^{5}\right) .$ Here, $\vec{m}=m\hat{e}_{x}$ is the
ferromagnetic particle's magnetic moment, $\vec{n}$ is the unit vector
pointing in the direction from the tip to the spin, and $\vec{r}$ in the $yz$%
-plane is the equilibrium position of the tip. We assume that the spins are
confined in a very small volume. In the gradient $G$, $r_{z}=\vec{r}\cdot
\hat{e}_{z}$. Besides the magnetic field $B\left( z\right) $ produced by the
magnetized tip, the spins are also exposed to two static magnetic fields $%
\vec{B}_{s}=\left( -A,0,0\right) $, and $\vec{B}_{0}$.

The magnetized tip is described as a harmonic oscillator with the effective
mass $M$ and frequency $\omega $. With a probe microwave field $\vec{B}_{%
\mathrm{p}}=-\hat{e}_{x}B_{\mathrm{p}}\cos \Omega t,$ the Hamiltonian $H$\
of this spin-NAMR hybrid system is $H=H_{0}+H_{I}$, where%
\begin{equation}
H_{0}=\frac{p^{2}}{2M}+\frac{1}{2}m\omega ^{2}z^{2}+\sum_{j=1}^{N}\left(
g\sigma _{j}^{x}z+g_{0}\sigma _{j}^{z}\right) .  \label{h}
\end{equation}%
with $p$ the momentum of the NAMR. Here, the spin-NAMR coupling strength $%
g=g_{s}\mu _{B}G/2$.

Actually, when $N$ is large and with the low excitations of the spins,
following the similar procedure to that in the last section, the Hamiltonian
in Eq. (\ref{h}) is rewritten as%
\begin{eqnarray}
H_{0} &=&\frac{\hbar \omega _{0}}{2}\left( P_{0}^{2}+Z_{0}^{2}\right) +\frac{%
\hbar \omega }{2}\left( P^{2}+Z^{2}\right)   \notag \\
&&+\hbar G\sqrt{N}Z_{0}Z,  \label{h1}
\end{eqnarray}%
where%
\begin{equation}
Z=\sqrt{\frac{M\omega }{\hbar }}z,P=\frac{p}{\sqrt{\hbar M\omega }},
\end{equation}%
and the NAMR-spin ensemble coupling constant is $G=g\sqrt{2\hbar /m\omega }%
/\hbar $.

It is shown in Eq. (\ref{h1}) that under the low excitation approximation
with large $N$ limit, the NAMR-spin ensemble coupling system is described by
a two harmonic coupling system if $\omega _{0}>0$, with the coupling
constant proportional to $\sqrt{N}$. In large $N$ limit with low
excitations, $H_{I}$ is written as%
\begin{equation}
H_{I}=\hbar \sqrt{N}G_{\mathrm{p}}Z_{0}\left( e^{-i\Omega t}+e^{i\Omega
t}\right) ,  \label{hi}
\end{equation}%
where $G_{\mathrm{p}}=g_{s}\mu _{B}B_{\mathrm{p}}/\left( \sqrt{2}\hbar
\right) $. The set of Heisenberg-Langevin equations gives%
\begin{eqnarray}
\partial _{t}^{2}Z_{0} &=&-\gamma _{0}\dot{Z}_{0}-\omega
_{0}^{2}Z_{0}-\omega _{0}\sqrt{N}GZ  \notag \\
&&-\omega _{0}\sqrt{N}G_{\mathrm{p}}\left( e^{-i\Omega t}+e^{-i\Omega
t}\right) ,  \label{a1}
\end{eqnarray}%
\begin{equation}
\partial _{t}^{2}Z=-\gamma \dot{Z}-\omega ^{2}Z-\omega \sqrt{N}GZ_{0},
\label{a2}
\end{equation}%
where $\gamma _{0}$ ($\gamma $) is the decay rate for $Z_{0}$ ($Z$). The
probe microwave field also provides a \textquotedblleft driving" term in the
set of equations (\ref{a1}) and (\ref{a2}), as what the probe light behaves
in the conventional EIT phenomenon. Here, we have ignored the fluctuations
as we are interested in the steady states and the fluctuations' expectation
values on the steady states are zero. The solutions to Eqs. (\ref{a1}) and (%
\ref{a2}) have the form%
\begin{equation}
Z_{0s}\left( t\right) =Z_{0s}\left( \Omega \right) e^{-i\Omega
t}+Z_{0s}\left( -\Omega \right) e^{i\Omega t},
\end{equation}%
and%
\begin{equation}
Z_{s}\left( t\right) =Z_{s}\left( \Omega \right) e^{-i\Omega t}+Z_{s}\left(
-\Omega \right) e^{i\Omega t}.  \label{a4}
\end{equation}%
It follows from Eqs. (\ref{a1})-(\ref{a4}) that the solution for $%
Z_{0s}\left( \Omega \right) $ is%
\begin{equation}
Z_{0}\left( \Omega \right) =\frac{\omega _{0}\sqrt{N}G_{\mathrm{p}}\xi }{%
-N\omega _{0}\omega G^{2}+\xi _{0}\xi },
\end{equation}%
where%
\begin{equation}
\xi _{0}=i\Omega \gamma _{0}-\omega _{0}^{2}+\Omega ^{2},
\end{equation}%
and%
\begin{equation}
\xi =i\Omega \gamma -\omega ^{2}+\Omega ^{2}.
\end{equation}%
The magnetic susceptibility of the alternating magnetic field $\vec{B}_{%
\mathrm{p}}$, $\chi _{M}$ is%
\begin{equation}
\chi _{M}=\frac{\vec{M}}{\vec{B}_{\mathrm{p}}/\mu _{0}-\vec{M}}\approx \frac{%
\mu _{0}\vec{M}}{\vec{B}_{\mathrm{p}}},  \label{exp}
\end{equation}%
where $\mu _{0}$ is the permeability of vacuum, and the magnetization
intensity $\vec{M}$ is%
\begin{eqnarray}
\vec{M} &=&\hat{e}_{x}\frac{g_{s}\mu _{B}}{2}\left\langle
\sum_{j=1}^{N}\sigma _{j}^{x}\right\rangle /V  \notag \\
&=&\hat{e}_{x}\frac{\sqrt{N}g_{s}\mu _{B}}{\sqrt{2}V}\left[ Z_{0s}\left(
\Omega \right) e^{-\mathrm{i}\Omega t}+\mathrm{c.c.}\right] ,
\end{eqnarray}%
with the volume of the spin ensemble $V$. Here, we have assumed that the
magnetization intensity $\left\vert \vec{M}\right\vert $ is small compared
with $\left\vert \vec{B}_{\mathrm{p}}\right\vert /\mu _{0}$, in order to
ensure the validity of the expansion in Eq. (\ref{exp}). Consequently, the
magnetic susceptibility $\chi _{M}\left( \Omega \right) $ is%
\begin{equation}
\chi _{M}\left( \Omega \right) =-\frac{\mu _{0}g_{s}\mu _{B}}{\sqrt{2}VB_{%
\mathrm{p}}}\sqrt{N}Z_{0s}\left( \Omega \right) .  \label{ee}
\end{equation}%
The real part and the imaginary part of $\chi _{M}\left( \Omega \right) $
depict the dispersive response and the absorption respectively. With the
parameters as (in the unit of $\omega _{0}$) $\omega =1$, $\gamma
_{0}=5\times 10^{-2}$, $\gamma =10^{-7}$, $G_{\mathrm{p}}=1$, $N=20$, $B_{%
\mathrm{p}}=\sqrt{2}\hbar G_{\mathrm{p}}/g_{s}\mu _{B}$, and $V=\left( 4\pi
/3\right) 10^{3}\mathrm{nm}^{3}$, we plot Re[$\chi _{M}\left( \Omega \right)
$] and Im[$\chi _{M}\left( \Omega \right) $] in Figs. 4(a) and 4(b), for $G=0
$ and $G=0.05$ respectively. In Fig. 4(a), the absorbed peak is at the
frequency $\Omega =\omega $, as the nuclear spin ensemble is decoupled with
the NAMR. The absorption window and slow light phenomenon for the microwave
field due to the coupling with the NAMR are illustrated in Fig. 4(b), which
shows the analog of EIT. We note that there are two absorption peaks in Fig.
4(b), corresponding approximately to the two eigen-frequencies derived from
Eq. (\ref{h1}). In the absorption window, the slope of Re[$\chi _{M}\left(
\Omega \right) $] is positive, which illustrates that the group velocity of
the microwave field is reduced dramatically.

\begin{figure}[tbp]
\includegraphics[bb=41 470 551 775, width=4 cm, clip]{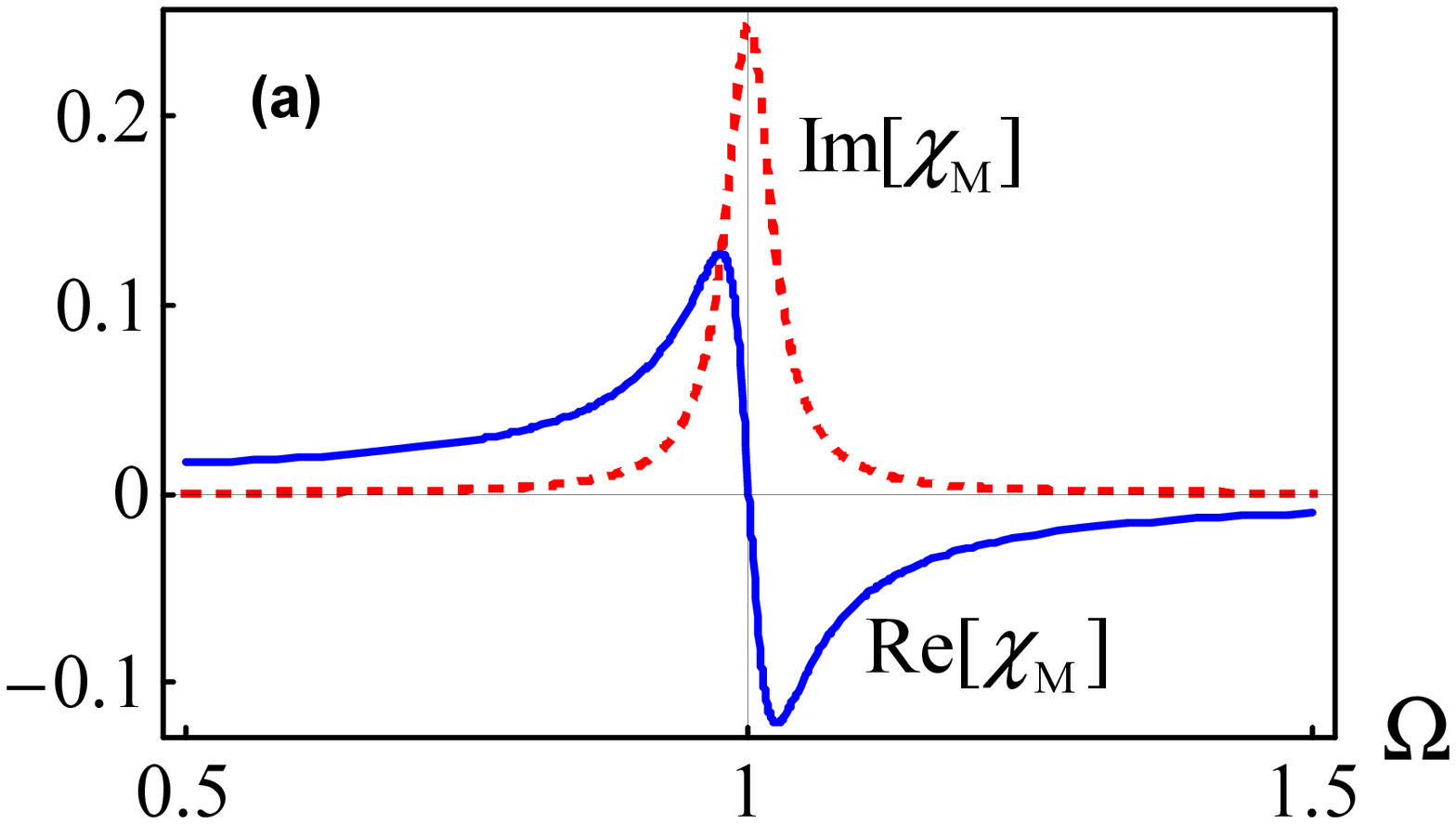} %
\includegraphics[bb=41 470 551 775, width=4 cm, clip]{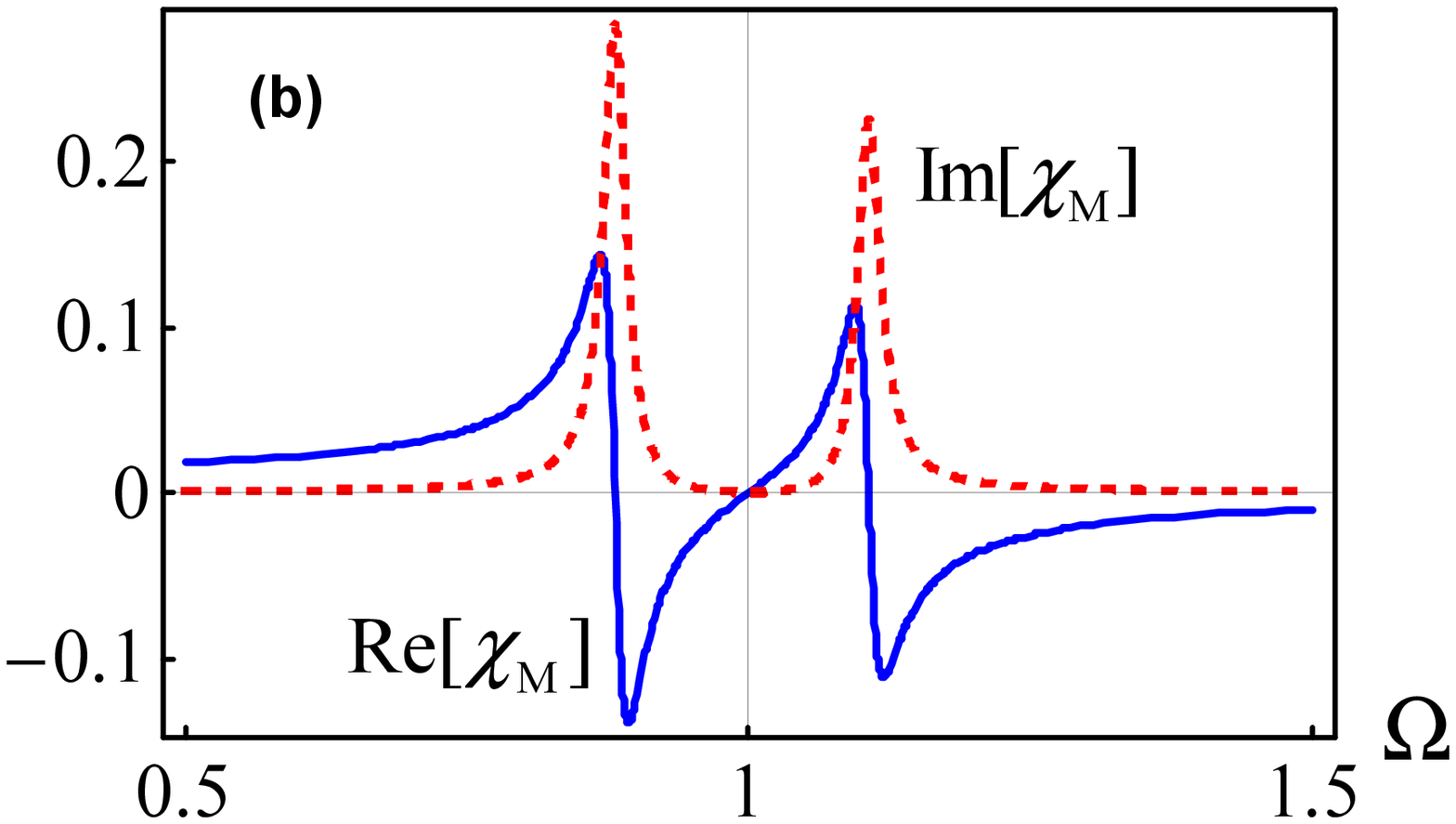}
\caption{(Color online) The frequency dependence of the real part
(the blue solid line) and the imaginary part (the red dashed line)
of the susceptibility $\protect\chi _{M}\left( \Omega \right) $ in
single NARM-spin ensemble coupling system. The NAMR-spin coupling
constant $G$ is: (a) $G=0$; (b) $G=0.05$. When The NAMR-spin
coupling exists, there is a window in the absorption spectrum, with
the positive slope of Re[$\protect\chi _{M}\left( \Omega \right) $]
in the window. This is an analog of EIT effect in the atomic
ensemble.}
\end{figure}

To see why the above mechanical system can display an EIT analog and its
intrinsic mechanism in detail, we revisit the EIT effect in an AMO system
shown in Fig. 3(b). Fig. 3(b) shows the energy levels of the $\Lambda $-type
atom of the atomic ensemble. Here, the single-mode driving field makes
transition between the excited state $\left\vert a\right\rangle $ and the
second lowest state $\left\vert c\right\rangle $ with the detuning $\Delta _{%
\mathrm{c}}=\omega _{\mathrm{ac}}-\nu _{\mathrm{c}}$, while the single-mode
probe light makes transition between the state $\left\vert a\right\rangle $
and the lowest state $\left\vert c\right\rangle $ with the detuning $\Delta
_{\mathrm{p}}=\omega _{\mathrm{ab}}-\nu _{\mathrm{p}}$. Here, $\omega _{%
\mathrm{ac}}$ ($\omega _{\mathrm{ab}}$) is the energy level spacing between
the states $\left\vert a\right\rangle $ and $\left\vert c\right\rangle $ ($%
\left\vert b\right\rangle $), and $\nu _{\mathrm{c}}$ ($\nu _{\mathrm{p}}$)
is the frequency of the driving (probe) light. In the rotating frame with
respect to~\cite{liyong1}%
\begin{equation}
\nu _{\mathrm{p}}S+\left( \omega _{\mathrm{ab}}-\omega _{\mathrm{ac}}\right)
\sum_{j=1}^{N_{a}}\left\vert c\right\rangle _{jj}\left\langle c\right\vert
+\nu _{\mathrm{p}}a^{\dag }a,
\end{equation}%
in the large $N_{a}$ (the number of atoms) limit with low excitations of the
atom ensemble, the Hamiltonian is%
\begin{equation}
H_{EIT}=\Delta _{\mathrm{p}}A^{\dag }A+\left( g_{\mathrm{p}}\sqrt{N_{a}}%
aA^{\dag }+g_{\mathrm{c}}e^{\mathrm{i}\left( \Delta _{\mathrm{c}}-\Delta _{%
\mathrm{p}}\right) t}A^{\dag }C+\mathrm{H.c.}\right) ,  \label{eit}
\end{equation}%
where the atomic collective excitation are described by%
\begin{equation}
A^{\dag }=\frac{1}{\sqrt{N_{a}}}\sum_{j=1}^{N_{a}}\left\vert a\right\rangle
_{jj}\left\langle b\right\vert ,\text{ }C=\frac{1}{\sqrt{N_{a}}}%
\sum_{j=1}^{N_{a}}\left\vert b\right\rangle _{jj}\left\langle c\right\vert ,
\label{e1}
\end{equation}%
and the operators defined in Eqs. (\ref{e1}) satisfy the commutation
relations approximately as \cite{liyong1} $\left[ A,A^{\dag }\right] \approx
1,$ $\left[ C,A^{\dag }\right] \approx 0,$ and $\left[ C,C^{\dag }\right]
\approx 1.$ Here, $a$ ($a^{\dag }$) is the annihilation (creation) operator
of the probe light, and $\left\vert \alpha \right\rangle _{jj}\left\langle
\beta \right\vert $ ($\alpha $, $\beta =a,b,c$) is $j$th atom's flip
operator. $g_{\mathrm{p}}$ ($g_{\mathrm{c}}$) is the coupling constant of
the probe (driving) light and a single atom with the corresponding energy
levels. We assume that both $g_{\mathrm{p}}$ and $g_{\mathrm{c}}$ are real.
It is shown in Eq. (\ref{eit}) that the EIT effect based on the $\Lambda $%
-type three level atomic ensemble can be re-explained by the coupling of two
\textquotedblleft harmonic oscillators" (depicted by the collective
excitation operators $A$ and $C$), with the coupling strength $g_{\mathrm{c}}
$. Here, the coupling of $A$-mode to the quantized field of $a$ can compare
with the semi-classical coupling in Eq. (\ref{h1}). Note that under the
rotating-wave approximation, the Hamiltonian $H$ in Eqs. (\ref{h1}) and (\ref%
{hi}) has the same form as $H_{EIT}$ in Eq. (\ref{eit}). As a result, the
hybrid system consisting of a NAMR and a nuclear spin ensemble can exhibit
the analog of EIT phenomenon.

\section{Double-EIT Analog and Slowing Light}

We have studied the analog of EIT effect in the last section for the basic
part of our hybrid NAMR-spin coupling network. In this section, we study the
double-EIT effect in the system consisting of two NAMRs coupled to a $N$
spin ensemble. We first rewritten the Hamiltonian $H_{0}^{d}$ as%
\begin{eqnarray}
H_{0}^{d} &=&\frac{\hbar \omega _{1}}{2}\left( P_{1}^{2}+Z_{1}^{2}\right) +%
\frac{\hbar \omega _{2}}{2}\left( P_{2}^{2}+Z_{2}^{2}\right)   \notag \\
&&+\frac{\hbar \omega _{0}}{2}\left( P_{0}^{2}+Z_{0}^{2}\right) +\hbar \sqrt{%
N}\sum_{j=1}^{2}G_{j}Z_{0}Z_{j}.  \label{1}
\end{eqnarray}%
Eq.~(\ref{1}) shows a coupled-oscillator system, where two harmonic
oscillators (NAMRs) couple to another oscillator (spin ensemble) with the
coupling constants strengthen by $\sqrt{N}$ respectively. The interaction of
the spin ensemble and the probe microwave field is described in Eq. (\ref{hi}%
).

With the same procedure as that in the last section, the steady state
solution $Z_{0}^{d}\left( \Omega \right) =\omega _{0}\sqrt{N}G_{\mathrm{p}%
}\xi _{1}\xi _{2}/D\left( \Omega \right) $, where

\begin{equation}
D\left( \Omega \right) =-N\omega _{0}\left( \omega _{2}G_{2}^{2}\xi
_{1}-\omega _{1}G_{1}^{2}\xi _{2}\right) +\xi _{0}\xi _{1}\xi _{2},
\end{equation}%
and%
\begin{equation}
\xi _{j}=i\Omega \gamma _{j}-\omega _{j}^{2}+\Omega ^{2},j=0,1,2.
\end{equation}%
Here, $\gamma _{j}$ ($j=1,2$) is the decay rate of the $j$th NAMR.

Consequently, the magnetic susceptibility $\chi _{M}^{d}$ is%
\begin{equation}
\chi _{M}^{d}\left( \Omega \right) =-\frac{\mu _{0}g_{s}\mu _{B}}{\sqrt{2}%
VB_{\mathrm{p}}}\sqrt{N}Z_{0}^{d}\left( \Omega \right) ,
\end{equation}%
whose real part and imaginary part depict the dispersive response and the
absorption respectively. We note that, generally, when the decay rates $%
\gamma _{0}\ll \omega _{0}$, $\gamma _{1}\ll \omega _{1}$, and $\gamma
_{2}\ll \omega _{2}$, $D\left( \Omega \right) $ is approximately zero with 3
non-negative real values of $\Omega $, which means that there are three
absorbing peaks in $\chi _{M}^{d}\left( \Omega \right) $. Actually, we can
also observe the three absorbed peaks without referring to the steady state
solution $Z_{0}^{d}\left( \Omega \right) $. From the Hamiltonian (\ref{1}),
the Heisenberg equations follow as%
\begin{eqnarray}
&&\left( -\omega _{0}^{2}+\Omega ^{2}\right) Z_{0}\left( 0\right) -\omega
_{0}\sqrt{N}\left[ G_{1}Z_{1}\left( 0\right) +G_{2}Z_{2}\left( 0\right) %
\right]   \notag \\
&=&\omega _{0}\sqrt{N}G,
\end{eqnarray}%
\begin{equation}
\left( -\omega _{1}^{2}+\Omega ^{2}\right) Z_{1}\left( 0\right) -\omega _{1}%
\sqrt{N}G_{1}Z_{0}\left( 0\right) =0,
\end{equation}%
\begin{equation}
\left( -\omega _{2}^{2}+\Omega ^{2}\right) Z_{2}\left( 0\right) -\omega _{2}%
\sqrt{N}G_{2}Z_{0}\left( 0\right) =0.
\end{equation}%
Obviously, the determinant%
\begin{equation}
\det \left(
\begin{array}{ccc}
-\omega _{0}^{2}+\Omega ^{2} & -\omega _{0}\sqrt{N}G_{1} & -\omega _{0}\sqrt{%
N}G_{2} \\
-\omega _{1}\sqrt{N}G_{1} & -\omega _{1}^{2}+\Omega ^{2} & 0 \\
-\omega _{2}\sqrt{N}G_{2} & 0 & -\omega _{2}^{2}+\Omega ^{2}%
\end{array}%
\right)   \label{e}
\end{equation}%
is just $D\left( \Omega \right) $. Thus, the vanishing determinant means the
three peaks correspond to the three eigen-frequencies in the Hamiltonian (%
\ref{1}). This is the physical mechanism of the mechanical analog of double
EIT effect.

In Figs. 5(a)-5(d), we plot the real part and the imaginary part of $\chi
_{M}^{d}\left( \Omega \right) $ versus the microwave field's frequency $%
\Omega $ with different values of $G_{1}$ and $G_{2}$, while other
parameters are fixed as (in the unit of $\omega _{0}$) $\omega _{1}=1$, $%
\omega _{2}=1.5$, $\gamma _{0}=5\ast 10^{-2}$, $\gamma _{1}=\gamma
_{2}=10^{-7}$, $G=1$, $N=20$, $B=\sqrt{2}\hbar G/g_{s}\mu _{B}$, and $%
V=\left( 4\pi /3\right) 10^{3}\mathrm{nm}^{3}$. It is shown in Fig. 5(a)
that when the coupling strength $G_{1}=G_{2}=0$, the single absorbed peak
appears at the frequency $\omega _{0}$. When we increase $G_{1}$ and $G_{2}$%
, there are three absorbed peaks with two windows, each of which is
localized between the nearest two absorption peaks. Figs. 5(b)-5(d)
illustrate the double EIT effect with three peaks corresponding to three
non-degenerate solutions to the equation $D\left( \Omega \right) =0$. We
notice that in some situations, the absorption peaks degenerate to two even
if the solutions to $D\left( \Omega \right) =0$ are non-degenerate. For
example, when $\omega _{1}\approx \omega _{2}$, which leads to $\xi
_{1}\approx \xi _{2}=\xi $, the magnetic susceptibility $\chi _{M}^{d}\left(
\Omega \right) $ becomes

\begin{equation}
\chi _{M}^{d}\left( \Omega \right) \approx \frac{N\mu _{0}g_{s}\mu
_{B}\omega _{0}G_{\mathrm{p}}\xi }{\sqrt{2}VB_{\mathrm{p}}\left[ N\omega
_{0}\left( \omega _{2}G_{2}^{2}-\omega _{1}G_{1}^{2}\right) -\xi _{0}\xi %
\right] }.  \label{p1}
\end{equation}%
There are only two non-negative roots for the zeroes of the dominator in the
right hand side of Eq. (\ref{p1}), corresponding to two resonant peaks in
the absorbing spectrum. This situation is illustrated in Fig. 6, with the
same parameters as that in Fig. 5(b), except for the NAMRs' frequencies $%
\omega _{1}=\omega _{2}=1$.

\begin{figure}[tbp]
\includegraphics[bb=41 470 551 775, width=4 cm, clip]{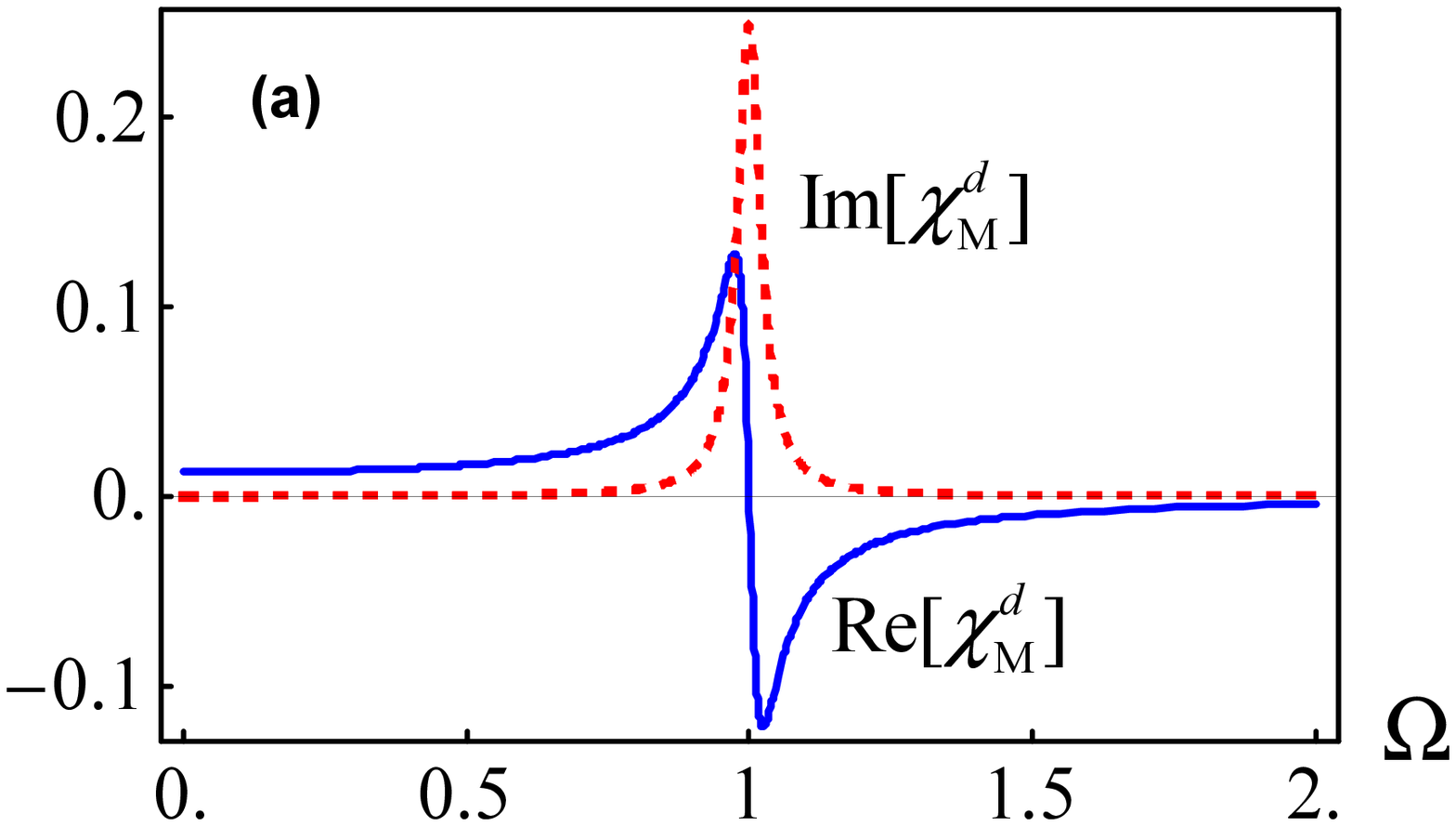} %
\includegraphics[bb=41 470 551 775, width=4 cm, clip]{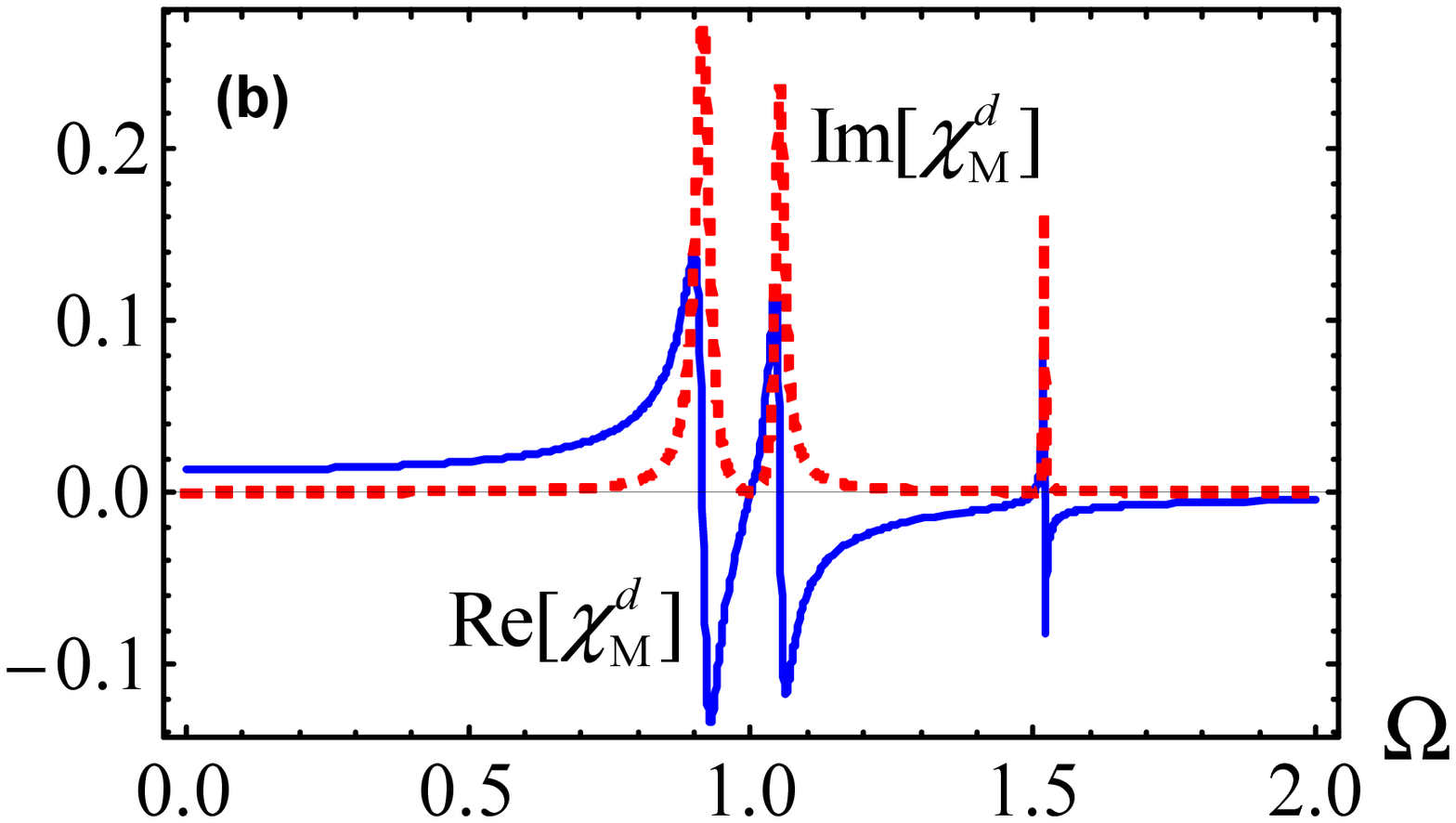} %
\includegraphics[bb=41 470 551 775, width=4 cm, clip]{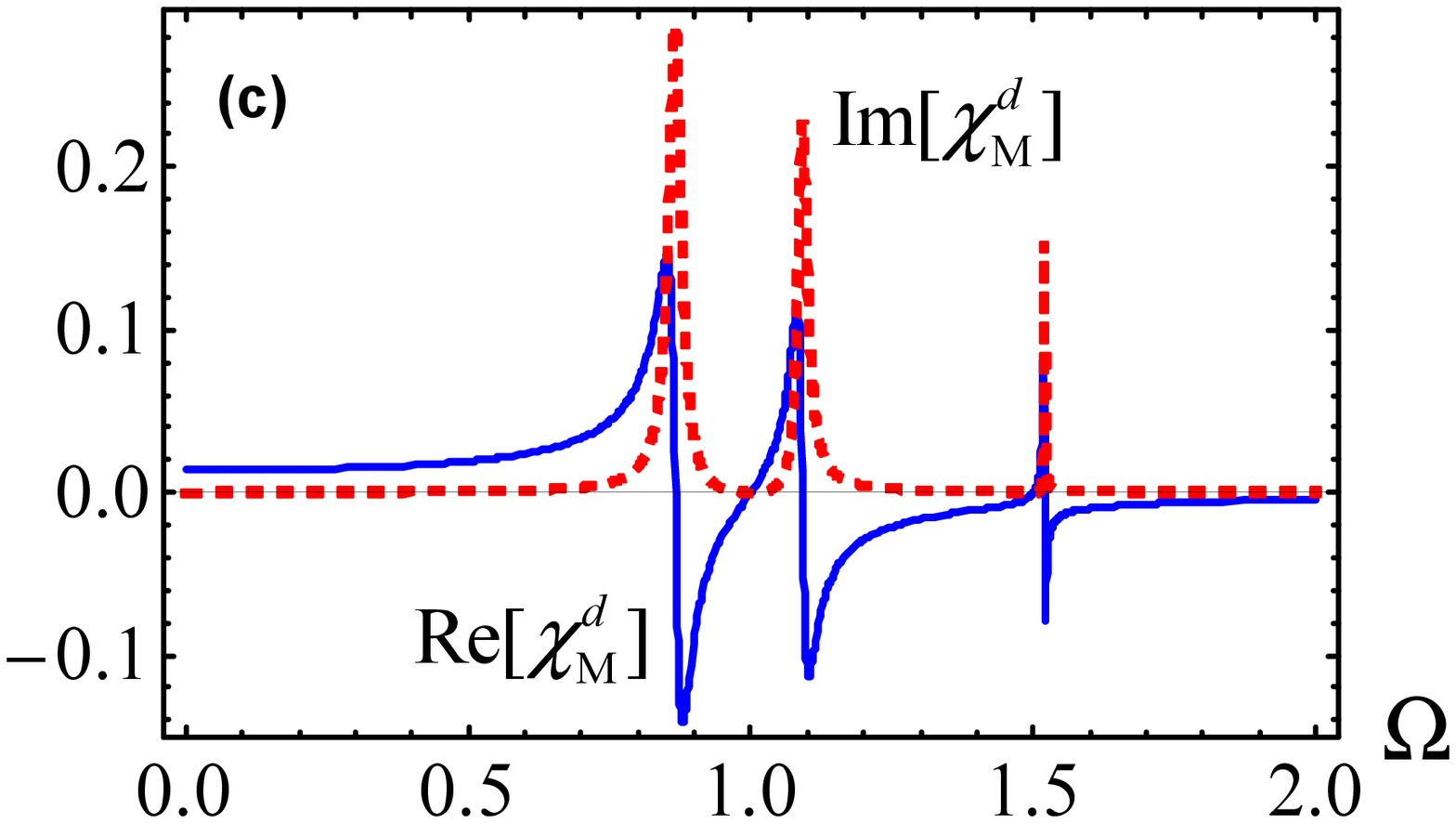} %
\includegraphics[bb=41 470 551 775, width=4 cm, clip]{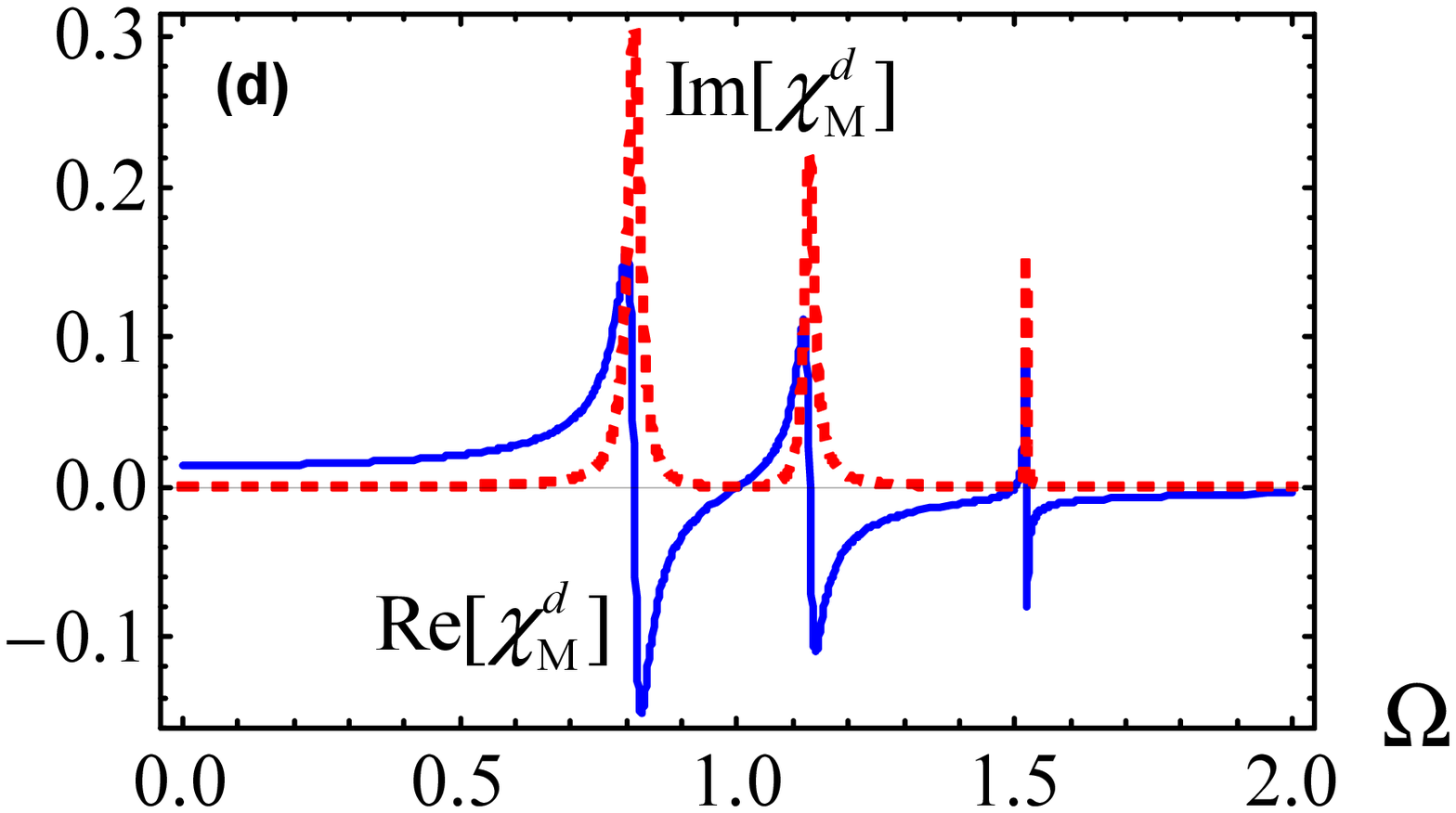}
\caption{(Color online) The frequency dependence of the real part (the blue
solid line) and the imaginary part (the red dashed line) of the magnetic
susceptibility $\protect\chi _{M}\left( \Omega \right)$. The values of $%
G_{1} $ and $G_{2}$ are: (a)$G_{1}= G_{2}=0$; (b)$G_{1}=0.03$ and $%
G_{2}=0.05 $; (c)$G_{1}=G_{2}=0.05$; (d)$G_{1}=0.07$ and $G_{2}=0.05$. When $%
G_{1}$ and $G_{2}$ are not zero, the double-EIT effect appears with two
absorption windows.}
\end{figure}

\begin{figure}[tbp]
\includegraphics[bb=41 470 551 775, width=7 cm, clip]{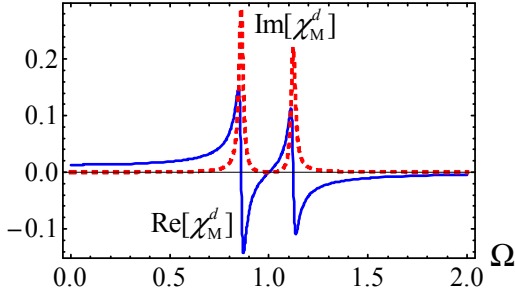}
\caption{(Color online) The frequency dependence of the real part (the blue
solid line) and the imaginary part (the red dashed line) of the magnetic
susceptibility $\protect\chi _{M}^{d}\left( \Omega \right)$ in some special
situation, where the two absorption windows are reduced to one.}
\end{figure}

Finally, to witness the existence of the double-EIT phenomenon in our setup,
we consider the velocity of signal transfer as follows. The group velocity
of the alternating magnetic field propagating in the spin ensemble is
defined as \cite{liyong1}%
\begin{eqnarray}
v_{g} &=&\text{Re}\left[ \frac{\mathrm{d}\Omega }{\mathrm{d}\left[ \Omega
n\left( \Omega \right) /c\right] }\right] \\
&=&\text{Re}\left[ \frac{c}{n\left( \Omega \right) +\Omega \partial _{\Omega
}n\left( \Omega \right) }\right] ,
\end{eqnarray}%
where $n\left( \Omega \right) $ is the complex refractive index defined as%
\begin{equation}
n\left( \Omega \right) =\sqrt{1+\chi _{M}\left( \Omega \right) },
\end{equation}%
and $c$ is the velocity of light in vacuum. The group velocity (in unit of
the light velocity $c=1/\sqrt{\varepsilon _{0}\mu _{0}}$ in vacuum) in
frequency region between the first and last two absorbed peaks, is
illustrated in Fig. 7(a) and 7(b) respectively, with the parameters the same
as that in Fig. 5(b). It is shown in Fig. 7 that in both of the two
absorption windows, the group velocity of the microwave field is reduce
dramatically. It is indeed similar to that in the atomic EIT effect.

\begin{figure}[tbp]
\includegraphics[bb=56 442 558 774, width=7 cm, clip]{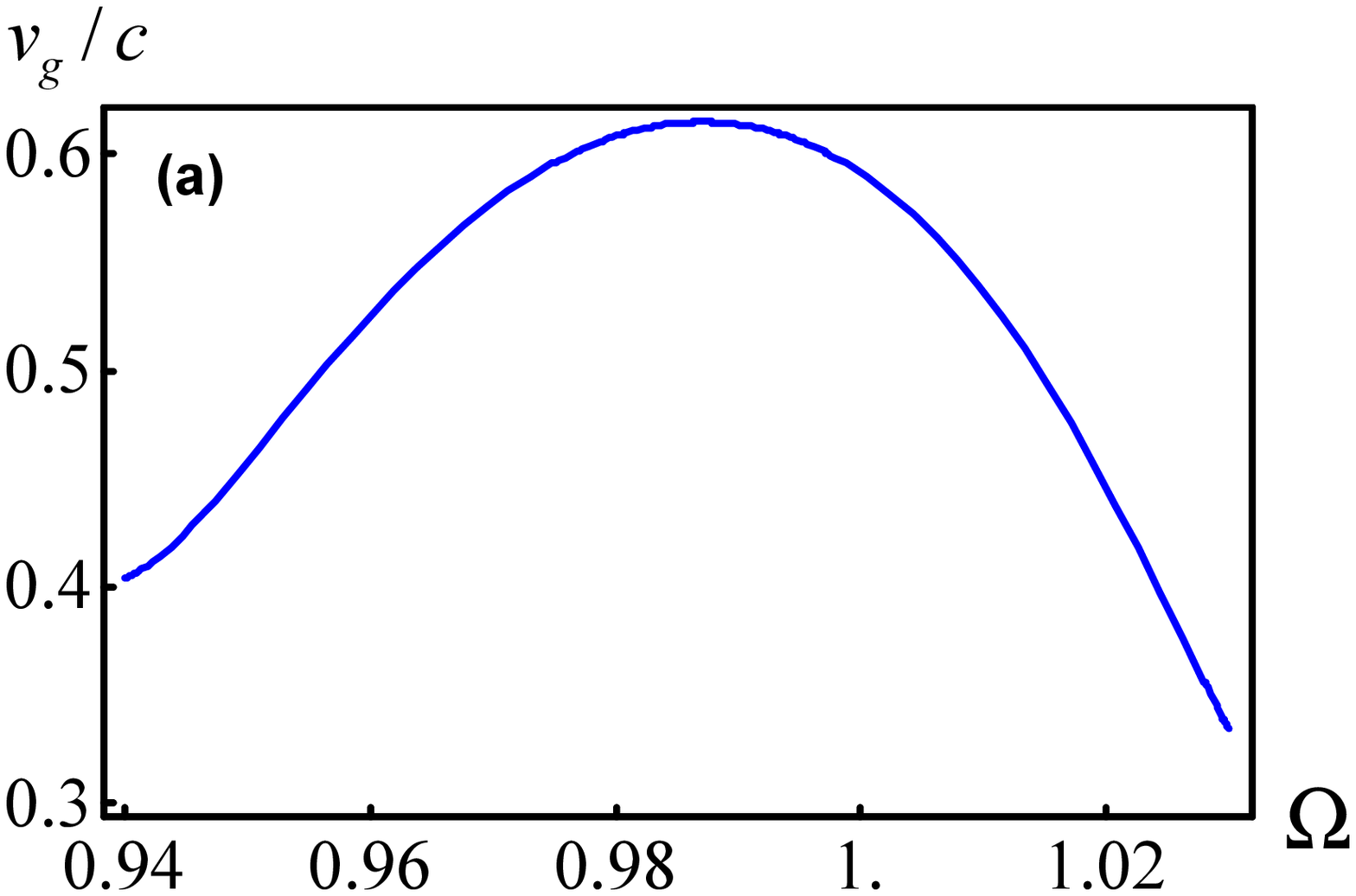} %
\includegraphics[bb=56 442 558 774, width=7 cm, clip]{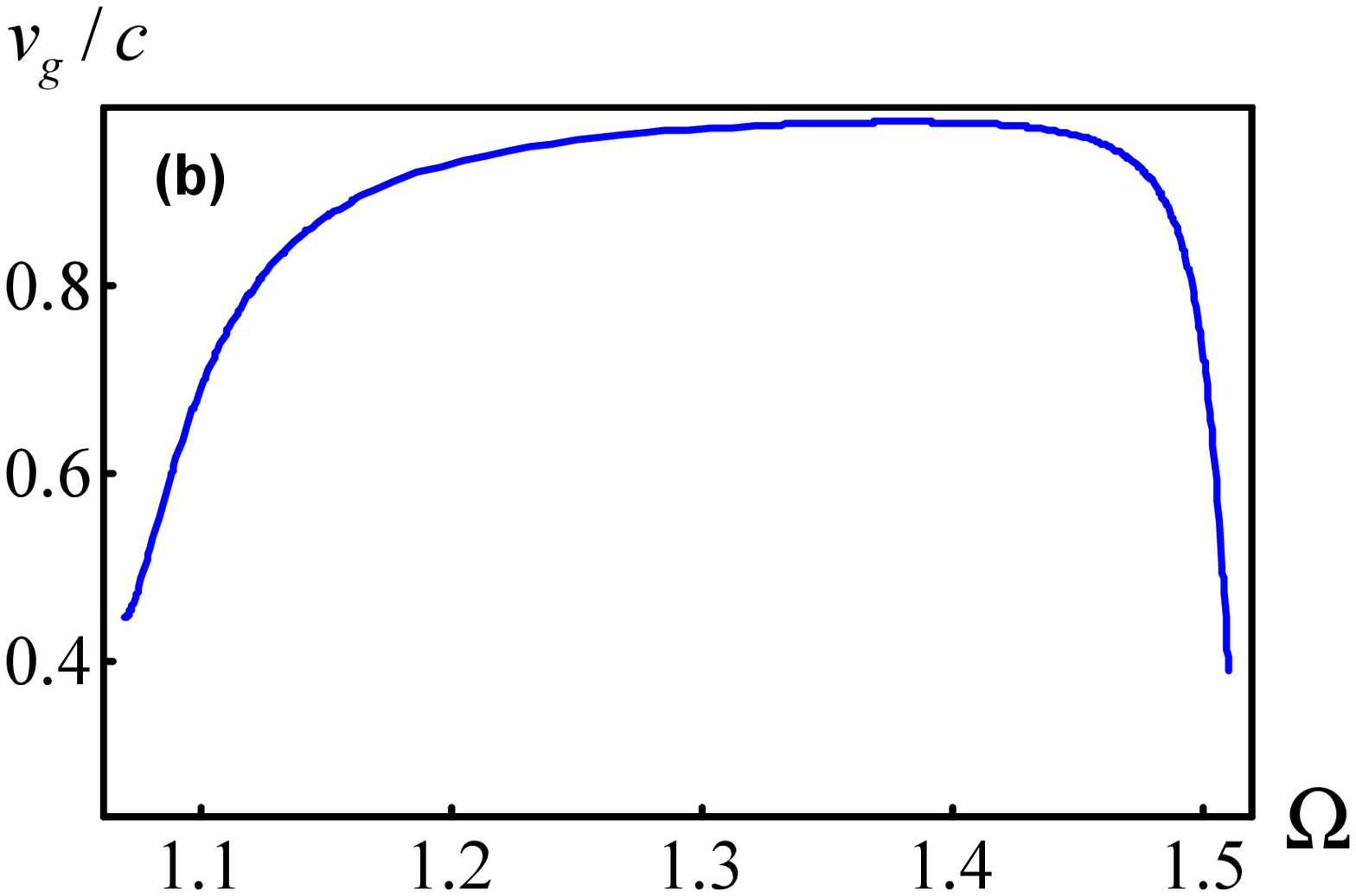}
\caption{The group velocity $v_{g}$ in the frequency region between
the first two [for (a)] and the last two [for (b)] absorbed peaks in
Fig. 5(b). The microwave field's group velocity is reduced
dramatically in both of these two windows.}
\end{figure}

\section{Summary}

We have proposed and studied a hybrid setup where two NAMRs are coupled to a
nuclear spin ensemble to demonstrate quantum interference phenomenon, i.e.,
an analog of EIT in atomic ensemble coupled to light. This system is
implemented by cantilevers tipped with ferromagnetic particles producing
inhomogeneous magnetic fields which couple the mechanical tips to the spin
ensemble. We have studied the dynamical properties in this NAMR-spin
ensemble-NAMR system by applying a probe microwave field. In the low
excitation approximation with large $N$ limit, this NAMR-spin ensemble-NAMR
coupling system behaves as a system of three coupled harmonic oscillators.
As a result, there is the so-called double-EIT effect in this system with
two absorption windows. Furthermore, we have shown the group velocity of the
microwave field is reduced dramatically in both of these two windows.

Finally, we point out that the NAMR-spin ensemble-NAMR coupling system is a
sub-network of such a structure consisting of an array of NAMRs and nuclear
spin ensembles, where the quantum information of the NAMR can be stored in
the nuclear spin ensemble for long time and transferred to the next NAMR in
a distance. And this process is repeated in the next sub-networks.
Therefore, it is expected that the spin ensembles can behave as a quantum
transducer that stores and transfer quantum information of the NAMRs.

\acknowledgments The work is supported by National Natural Science
Foundation of China a under Grant Nos. 10935010 and 11074261.%

\end{document}